\documentclass[pra,twocolumn,notitlepage,superscriptaddress]{revtex4-1}

\PassOptionsToPackage{usenames,dvipsnames}{color}
\usepackage{amsmath,amsfonts,amssymb,amsthm,graphicx,bbm,enumerate,times}
\usepackage{mathtools}
\usepackage{import}
\usepackage[normalem]{ulem}
\usepackage{hhline}
\usepackage{hyperref}
\usepackage{enumitem}
\usepackage{svg}
\usepackage{tikz}
\usetikzlibrary{shapes.geometric}
\usetikzlibrary{shapes.symbols}
\usepackage{wasysym}
\usepackage{makecell}
\usepackage{lipsum}
\newcommand{\B}{\mc B}
\newcommand{\sC}{B}
\newcommand{\sS}{A}
\renewcommand{\H}{\mc H}
\newcommand\M{\mc M}

\definecolor{RedOrange}{rgb}{1.0, 0.2, 0.0}

\newcommand{\ox}{\otimes}

 \newtheorem{theorem}{Theorem}
 \newtheorem{proposition}[theorem]{Proposition}

 \newtheorem{lemma}[theorem]{Lemma}
 
 \newtheorem{definition}[theorem]{Definition}

\definecolor{DarkGreen}{rgb}{0.5, 0.5, 0}
\definecolor{teal}{rgb}{0.0, 0.5, 0.5}
\newcommand{\mc}[1]{\mathcal{#1}}

\newcommand{\mb}[1]{\mathbb{#1}}

\newcommand{\tr}{\mathrm{Tr}} 
\newcommand{\Tr}{\mathrm{Tr}} 

\newcommand{\1}{\mathbf{1}}

\newcommand{\CC}{\mb{C}}
\newcommand{\RR}{\mb{R}}

\newcommand{\ket}[1]{\left.\left|{#1}\right.\right\rangle}

\newcommand{\bra}[1]{\left.\left\langle{#1}\right.\right|}

\newcommand{\ketbra}[2]{\ket{#1} \!\! \bra{#2}}

  \newcommand{\proj}[1]{\ketbra{#1}{#1}}

\newcommand{\hannover}{Leibniz Universit\"at Hannover, Appelstraße 2, 30167 Hannover, Germany}
\newcommand{\ntu}{School of Physical and Mathematical Sciences, Nanyang Technological University, 637371, Singapore}
\newcommand{\unist}{Department of Physics, Ulsan National Institute of Science and Technology (UNIST), Ulsan 44919, Republic of Korea}

\begin{document}
\title{Thermal operations from informational equilibrium}
\author{Seok Hyung Lie}
\email{seokhyung@unist.ac.kr}
\affiliation{\unist}
\author{Jeongrak Son}
\affiliation{\ntu}
\author{Paul Boes}
\noaffiliation
\author{Nelly H.Y. Ng}
\affiliation{\ntu}
\affiliation{Centre for Quantum Technologies, National University of Singapore, 3 Science Drive 2, 117543, Singapore}
\author{Henrik Wilming}
\email{henrik.wilming@itp.uni-hannover.de}
\affiliation{\hannover}
\begin{abstract}
    Thermal operations are quantum channels that have taken a prominent role in deriving fundamental thermodynamic limitations in quantum systems. We show that these channels are uniquely characterized by a purely quantum information theoretic property: They admit a dilation into a unitary process that leaves the environment invariant when applied to the equilibrium state. In other words, they are the only channels that preserve equilibrium between system and environment. Extending this perspective, we explore an information theoretic idealization of heat bath behavior, by considering channels where the environment remains locally invariant for every initial state of the system. These are known as catalytic channels. We show that catalytic channels provide a refined hierarchy of Gibbs-preserving maps for fully-degenerate Hamiltonians, and are closely related to dual unitary quantum circuits.
\end{abstract}
\maketitle
\textit{Introduction.---}%
Equilibrium is a central notion in physics, yet it manifests in different forms. 
First, there is an equilibrium equated with being stationary: A system is in equilibrium if its state remains invariant under the prescribed dynamics. 
This perspective applies primarily to individual systems.
Alternatively, equilibrium can be relational: multiple systems are said to be in equilibrium when they undergo an interaction, which leaves the local state of each system invariant. 
For example, thermodynamic equilibrium describes a situation in which the average flux of conserved quantities such as energy between two bodies is balanced, i.e. their macroscopic states remain unchanged.

Different notions of equilibrium naturally lead to different notions of \textit{equilibrating processes} through which physical systems approach equilibrium.
In the context of quantum thermodynamics, where equilibrium states are typically identified with Gibbs states, various proposals exist for equilibrating processes, modeled by sets of quantum channels. One well-known case is \emph{thermal operations} \cite{Horodecki2013, Brandao2013, Cwiklinksi2015}, based on the idea of equilibrium as a relational property with respect to heat baths. Another is Gibbs-preserving maps \cite{Faist2015MinWork, Faist2015GPvsTO, FaistRenner2018}, whose only constraint being that they preserve the Gibbs state, modeling equilibrium as a single-system stationarity. The two classes are known to be distinct \cite{Faist2015GPvsTO, Lostaglio2015, TajimaTakagi2025}, and fully understanding their difference has been one of the major problems of quantum thermodynamics. However, few attempts have been made to understand the gap by comparing different notions of equilibrium.

To pinpoint the origin of their difference, we begin with a simple, information theoretic notion of equilibrium in the context of quantum mechanics. We show that this notion naturally gives rise to a thermodynamic interpretation, allowing us to characterize thermal operations as processes that equilibrate information. Our results show that whenever a system interacts with an environment in a way such that the induced channel on the system preserves the Gibbs state but is not a thermal operation, then the environment \textit{must} change. This is true even if the system begins in a fixed point of that channel i.e., an equilibrium defined by single‐system stationarity.

\textit{Main results.---}%
While the notion of equilibrium is often associated with thermodynamics, it is possible to define equilibrium purely information theoretically without presupposing thermodynamical concepts. 
In a closed quantum system, dynamics (for a fixed time) is described by a unitary operator $U$. 
A state (density matrix) $\omega$ is said to be stationary with respect to $U$ if $U\omega U^\dagger = \omega$.
When the system consists of two subsystems $A$ and $B$, we can say that two local states $\omega_A$ and $\omega_B$ are in  (informational) equilibrium relative to $U$ if 
\begin{align}\label{eq:local-equilibrium}
	\tr_{B}(\sigma_{AB}) = \omega_A,\quad \tr_{A}(\sigma_{AB}) = \omega_B,
\end{align}
where $\sigma_{AB} = U\omega_A\ox \omega_B U^\dagger$ is the time-evolved state. We show (Lemma \ref{lemma:basic}) that this concept naturally generalizes to any number of subsystems.
In particular, if $(\omega_A,\omega_B)$ and $(\omega_B,\omega_C)$ are in equilibrium under $U_{AB}$ and $U_{BC}$, respectively, the triple $(\omega_A,\omega_B,\omega_C)$ is in equilibrium under an appropriate unitary dynamics on $ABC$. 
In this sense, equilibrium is transitive, just like thermodynamic equilibrium is transitive by the zeroth law of thermodynamics.

Now suppose that $\omega_A$ and $\omega_B$ are in equilibrium under $U$, but system $A$ is prepared in a state $\rho_A$ that may differ from $\omega_{A}$.
The effective dynamics on $A$ is described by 
\begin{align}\label{eq:dilation}
	T_A(\rho) = \tr_B(U \rho_A \ox \omega_B U^\dagger).
\end{align}
Since $(\omega_A,\omega_B)$ are in equilibrium, it follows that $\omega_{A}$ is a fixed-point of $T_A$, i.e. $T_A(\omega_A) = \omega_A$.
This implies that $T_A$ can only drive system $A$ closer to its equilibrium state $\omega_A$.
Hence, we can interpret $T_A$ as an open-system equilibration dynamics. 
We formally define such equilibration dynamics as follows. 
Throughout this work, we only consider finite-dimensional Hilbert spaces. 
\begin{definition}[Equilibrating dilation]\label{def:catalytic-FP}
	Consider a quantum channel $T_A$ on $A$, and a fixed-point $\omega_A$ of $T_A$. We say that $T_A$ has an \emph{equilibrating dilation} with respect to $\omega_A$ if there exists a dilation $(U,\omega_B)$ of $T_A$ (i.e. $T_A(\rho)$ is given as Eq.~\eqref{eq:dilation} for any $\rho$) such that
		\begin{align}\label{eq: ED equation}
			\tr_A(U \omega_A \ox \omega_B U^\dagger) = \omega_B.
		\end{align}
		The dilation is called \emph{non-degenerate} if $\omega_B$ has a non-degenerate spectrum. 
		A channel with an equilibrating dilation is called an equilibrating channel.
\end{definition} 

Not every quantum channel has an equilibrating dilation, and only when the channel has an equilibrating dilation $(U,\omega_B)$ the dynamics admit an equilibrium between $\omega_A$ and $\omega_B$.
Hence, in the first part of this work, we characterize the set of equilibrating channels, thereby identifying the effect of equilibration on each subsystem.
We will see that in the generic case of fixed-points of full rank,
after interpreting them as thermal equilibrium states,
this set precisely corresponds to the set of \emph{thermal operations} (Proposition \ref{prop:FP-dilations}), which has been widely used in quantum thermodynamics \cite{Janzing2000, Horodecki2013}.
In particular, this means that 
Eq.~\eqref{eq: ED equation}, encoding the equilibrium between system and environment,
excludes other well-studied classes of thermalization processes, such as enhanced thermal operations \cite{LostaglioEnTO, cwiklinski_limitations_2015} or Gibbs-preserving maps \cite{Faist2014}.

Another fundamental concept in classical thermodynamics related to equilibration is the idealized heat bath. 
Informally, a heat bath is a system whose state remains unchanged during a thermodynamic process, due to its vast size compared to any system it interacts with. Since interactions with finite systems involve only finite energy exchange, they cannot alter the bath’s temperature and thus its state. Such an information theoretic behavior of heat baths can be emulated by finite-sized quantum systems. By observing that the ability to equilibrate other systems without being altered matches the definition of a \emph{catalyst}, we now introduce the set of \emph{catalytic channels}.
\begin{definition}[Catalytic dilation]\label{def:catalytic}
		A quantum channel $T_{A}$ on $A$ has a \emph{catalytic dilation} $(U,\omega_B)$ if
		\begin{align}
			\tr_B(U\rho \ox\omega_B U^\dagger) &= T_{A}(\rho), \\ \tr_A(U \rho \ox \omega_B U^\dagger) &= \omega_B,
		\end{align}
		for all states $\rho$ on $A$. The dilation is called \emph{non-degenerate} if $\omega_B$ has a non-degenerate spectrum.
		A channel with a catalytic dilation is called a \emph{catalytic quantum channel}.
\end{definition}

\begin{figure}
    \centering
    \includegraphics[width=0.7\linewidth]{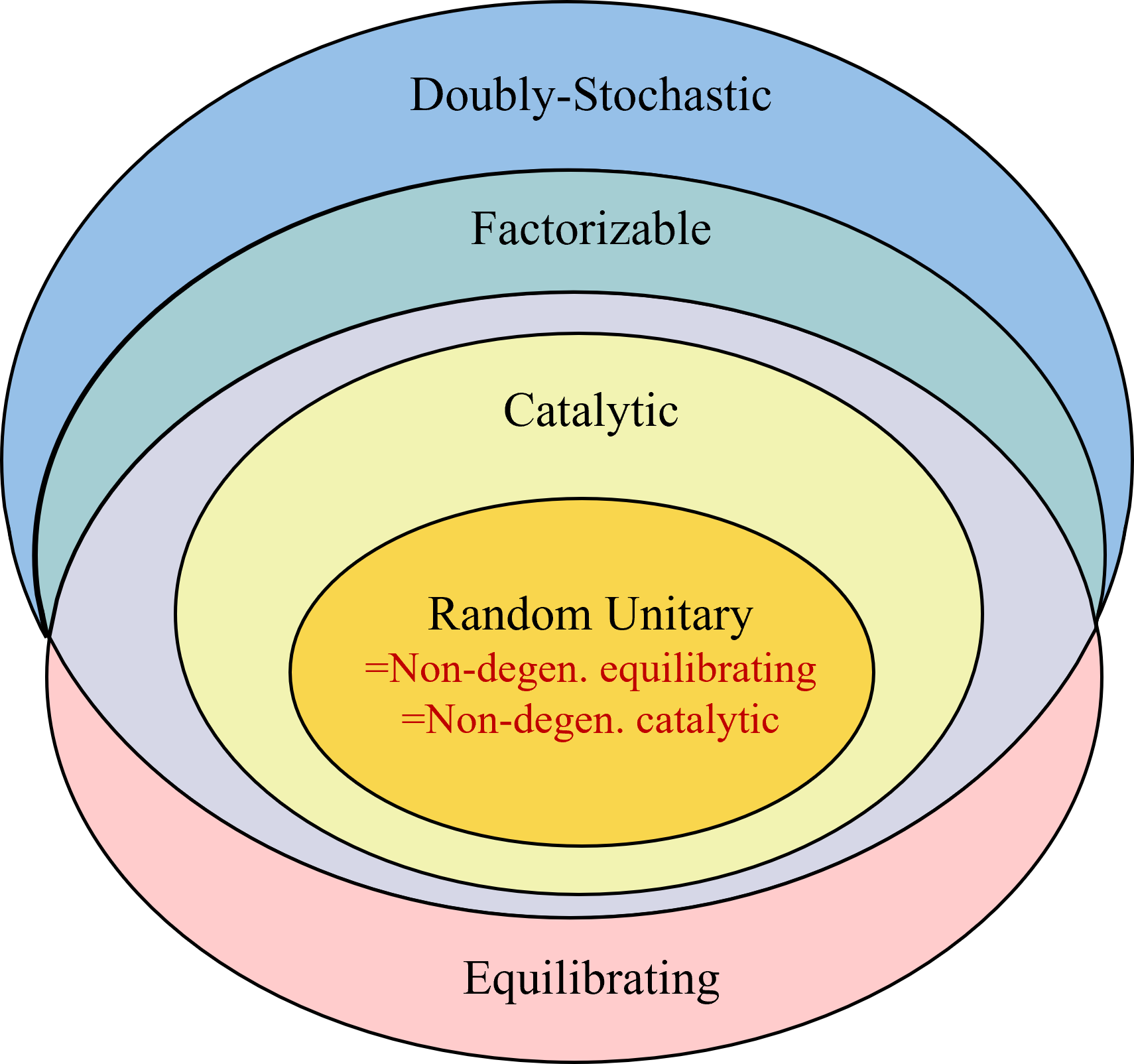}
	\caption{A hierarchy of doubly-stochastic channels, where all inclusions are strict. The strict inclusion of factorizable channels in doubly-stochastic channels is shown in \cite{haagerup_factorization_2010, haagerup_asymptotic_2015} and relies on the claim that Connes' embedding problem has a negative resolution \cite{ji_mipre_2022}.}
    \label{fig:hierarchy}
\end{figure}

The terms \emph{catalytic dilation} and \emph{catalytic quantum channel} draw inspiration from catalysis in chemistry: 
just as a chemical catalyst, the state of $B$ does not change, but its presence enables the implementation of non-unitary dynamics on $A$. 
See \cite{lipka-bartosik_catalysis_2024} for a comprehensive review on catalysis in quantum information theory. 
Catalytic channels have intriguing applications, including in quantum cryptography, where a message can be encoded in the correlation between $A$ and $B$ in a way that neither subsystem alone carries any information about the message~\cite{boes_catalytic_2018}.
This concept has been extended to the general advantage of having a catalytic access to the randomness of the auxiliary system $B$ \cite{boes_catalytic_2018, lie_randomness_2021, lie_catalytic_2021, lie2023_delocalized}, positioning the set of catalytic channels as an interesting class of its own.  
For instance, it is shown in \cite{lie_randomness_2021} that if $A$ is a finite-dimensional quantum system and $\omega_B$ has finite entropy, then the catalytic channel $T$ must be \emph{doubly-stochastic}, i.e. $T(1)=1$. 
This implies that $T$ can only \emph{increase} the entropy of $A$, which contrasts with the dynamics expected when a system is in contact with a finite temperature heat bath, where an increase in free energy does not necessarily accompany an increase in entropy.
In the context of thermodynamic analogy, the catalytic channel $T$ only mimics the dynamics resulting from an infinite temperature heat bath.

A simple subclass of catalytic channels are \emph{mixed-unitary channels}.
They have the form $T(\rho) = \sum_i p_i U_i\rho U_i^\dagger$ with some unitary operators $U_i$ and a probability distribution $(p_i)_i$, and admit a catalytic dilation $(U,\omega_{B})$ with
\begin{align}\label{eq: MU with controlled unitaries}
	U= \sum_i U_i\ox \proj{i},\quad \omega_B = \sum_i p_i \proj{i}.
\end{align} 
However, neither are all catalytic channels mixed unitaries 
(they are if and only if they allow a \emph{non-degenerate} catalytic dilation, see below), 
nor are all doubly-stochastic channels catalytic. In fact, we uncover a strict hierarchy of doubly-stochastic quantum channels using the notions of catalytic dilations and equilibrating dilations as well as \emph{factorizable maps}~\cite{Anantharaman-Delaroche2006,haagerup_factorization_2010}; see Fig.~\ref{fig:hierarchy}.

The emerging hierarchy stands in strong contrast to classical probability theory, where, by Birkhoff’s theorem \cite{birkhoff_three_1946}, a stochastic map is doubly-stochastic if and only if it is a mixed permutation. 
Catalytic dilations can also be defined for classical stochastic maps, in which case Birkhoff's theorem shows that a stochastic map admits a non-degenerate catalytic dilation if and only if it is a mixed permutation.
Identifying unitaries as the natural generalization of permutations, the quantum situation then precisely mirrors this latter statement, but not the former broader one about doubly-stochastic maps.

One may wonder how to classify catalytic channels. 
In \cite{lie_catalytic_2021} it was shown that these channels are induced by \emph{catalytic unitaries}, which are characterized by the property that their partial transpose $U^{\top_A}$ is still unitary. 
Catalytic unitaries are in correspondence with \emph{dual-unitary tensors}, a class of operators that has recently attracted significant attention for their role in constructing integrable quantum circuits \cite{piroli2020exact, foligno2024quantum, borsi2022construction}. 
The correspondence is shown in Appendix \ref{app:dual-unitary}.
Characterizing catalytic unitaries would therefore also provide a characterization of dual-unitary tensors. 
We leave this problem for future work.

\textit{Equilibration and thermal operations.---}
We elaborate on the emergence of thermodynamics from an information theoretic notion of equilibrium.
We begin by establishing the equivalence between global stationarity and equilibrium between subsystems.
\begin{lemma}[Informational zeroth law]\label{lemma:basic}
Suppose $\omega_A$ and $\omega_B$ are in equilibrium relative to $U$. Then
	\begin{align}
		U \omega_A \ox \omega_B U^\dagger = \omega_A \ox \omega_B.
	\end{align}
\end{lemma}
The first consequence of Lemma~\ref{lemma:basic} is that equilibrium is transitive, hence the name informational zeroth law. In other words, if $\omega_A$ and $\omega_B$ are in equilibrium relative to $U_{AB}$ and $\omega_B$ and $\omega_{C}$ relative to $U_{BC}$, the combined state $\omega_A\ox\omega_B\ox\omega_C$ are in equilibrium relative to any $U$ which is multiplicatively generated by  $(U_{AB}\ox 1_{C})^{t}$ and $(1_{A} \ox U_{BC})^{s}$ for $t,s\in\RR$.
Furthermore, Lemma~\ref{lemma: basic lemma multipartite} in End Matters generalizes Lemma~\ref{lemma:basic} for a generic unitary operator $U$ that preserves the marginal states of all subsystems.

The second consequence is that equilibrating channels can be characterized more easily, due to the unique designation of the final $AB$ composite state before the partial trace.
We use this together with the Lemma to show the following (see End Matter for proof):
\begin{proposition}\label{prop:FP-dilations}
		Let $T$ admit an equilibrating dilation $(U,\omega_B)$ with respect to $\omega_A$. Then
		\begin{enumerate}[leftmargin=*, itemsep=0pt]
			\item\label{item:FP-dilations1} $[U,\omega_A\ox\omega_B]=0$,
			\item\label{item:FP-dilations2} for every $t\in\RR$ and every state $\rho_A$ on $A$ we have $T(\omega^{i t}_A \rho_{A} \omega^{-it}_A) =\omega^{i t}_A T(\rho_{A}) \omega^{-it}_A$,
			\item\label{item:FP-dilations3} if $\omega_A$ has full rank, $\omega_B$ can be chosen to have full rank.
		\end{enumerate}
\end{proposition}

The third statement of Proposition~\ref{prop:FP-dilations} facilitates the thermodynamic interpretation of equilibrating channels.
When a system is in thermal equilibrium with a heat bath, its equilibrium state is the Gibbs state $e^{-\beta H}/Z$ corresponding to the system Hamiltonian $H$ and the bath temperature $\beta^{-1}=k_B T$. 
The Gibbs state always has full rank, and thus conversely any full-rank state $\omega$ is a Gibbs state corresponding to some Hamiltonian $\beta H + \log (Z)\1= -\log(\omega)$. 
Setting the ambient temperature $\beta^{-1}$  and the constant offsets $\log(Z_A)$ and $\log(Z_B)$ given in terms of $Z_A$ and $Z_B$ known as the partition function in thermodynamics for both $A$ and $B$, their Hamiltonians naturally emerge from the equilibrium states as $H_{A} \coloneqq -\beta^{-1}(\log(\omega_{A})+\log (Z_A))$ and $H_{B} \coloneqq -\beta^{-1}(\log(\omega_{B})+\log(Z_B))$.
If we arbitrarily fix the inverse temperature for the state $\omega_A$ on system $A$, this fixes the inverse temperature for all equilibrating dilations $(U,\omega_X)$ of equilibrating channels with $\omega_A$ as a fixed point. 

The second statement corroborates this thermodynamic interpretation. 
Taking $\omega_{A}$ to be the Gibbs state, we have $\omega_{A}^{it} = e^{-i\beta tH_{A}}/Z_A^{it}$. 
Then the channel $T$ is covariant under the time-translation symmetry generated by the Hamiltonian $H_{A}$~\cite{Keyl1999Cov, Marvian14_coherence}.
Although $H_{A}$ is not conserved during the evolution, since system $A$ is open to $B$, if one considers the closed dynamics of $AB$, the first statement of the theorem implies that the total Hamiltonian $H_A + H_B$ is conserved. 

Quantum channels of the form
\begin{align}
 T(\rho) = \tr_B\left(U \rho \ox \frac{e^{-\beta H_B}}{Z_B}U^\dagger\right)
\end{align}
with $[U, H_A + H_B]=0$ are known as \emph{thermal operations}~\cite{Janzing2000, Horodecki2013} and provide a coherent framework for resource-theoretic studies of thermodynamics~\cite{Ng18_Qthermo_book, Lostaglio19_review}. 
From Proposition~\ref{prop:FP-dilations}, thermal operations correspond precisely to quantum channels that admit an equilibrating dilation with respect to a full-rank fixed-point $\omega_{A} = {e^{-\beta H_A}}/{Z_A} $. 
This result aligns with the intuitive expectation that if a system starts in equilibrium, it not only remains stationary but also in equilibrium with the environment that remains in its Gibbs state.
Conversely this means that if $T$ is a quantum channel on $A$ that is supposed to model a thermal process on $A$ but is not a thermal operation, then \emph{for any dilation of the channel $T$, the environment must change its state, even when $A$ is already in a thermal equilibrium state.}
In other words, for every dilation $(U,\omega_B)$ of the channel $T$ we must have
\begin{align}\label{eq:A-on-B}
	\tr_A \left(U \frac{e^{-\beta H_A}}{Z_A} \ox \omega_B U^\dagger \right) \neq \omega_B.
\end{align}

Importantly, Eq.~\eqref{eq:A-on-B} implies that any such dilation \emph{must} contain non-equilibrium resources. 
This interpretation fits well with the recent findings that quantum channels with a thermal fixed-point -- but which are not thermal operations -- may require an infinite amount of coherence for their implementation \cite{tajima_gibbs-preserving_2024}. 
Our results also applies to enhanced thermal operations \cite{LostaglioEnTO, cwiklinski_limitations_2015}, i.e. channels with a thermal fixed-point and time-translation covariance.
Thanks to the time-translation covariance, they admit a dilation consisting of an energy-preserving $U$ satisfying $[U,H_{A}+H_{B}] = 0$, and an environment state $\omega_{B}$ such that $[\omega_{B},H_{B}] = 0$ \cite{Keyl1999Cov}. 
Hence, unlike general quantum channels with a thermal fixed-point, no coherence is needed for the implementation.
Yet, it is known that some enhanced thermal operations are not thermal operations~\cite{ding_exploring_2021}.
In this case, the LHS of Eq.~\eqref{eq:A-on-B} can be interpreted as describing a thermal operation $\rho \mapsto \tr_A (U \frac{e^{-\beta H_A}}{Z_A} \ox \rho U^\dagger )$ acting on $B$.
However, Eq.~\eqref{eq:A-on-B} indicates that $\omega_{B}$ cannot be a thermal state of $B$ at inverse temperature $\beta$.
It must therefore contain non-equilibrium free energy, which may be depleted after the operation. 

\textit{Hierarchy of doubly-stochastic channels.---}%
Having defined catalytic channels, we now examine the hierarchy of Gibbs-preserving maps at infinite temperature (i.e., trivial Hamiltonian), the so-called doubly-stochastic channels, and refine it by relating them to the class of catalytic channels.
Studying this hierarchy is important not only because it is a special case of more general Gibbs-preserving maps, but also because it may be possible to embed the structure for the nontrivial Hamiltonian case into that of the trivial one through the Gibbs-embedding map \cite{Brandao2015}, thereby enabling the translation of our understanding from the latter to the former.

As introduced earlier, catalytic channels imitate the behavior of idealized heat baths, which remain unchanged upon interacting with a finite system. 
We first discuss different ways to interpret and motivate catalytic channels.
From an information theoretic perspective, the fact that $B$ does not change irrespective of the input state on $A$, suggests that no information flows from $A$ to $B$. 
In quantum information theory, this is formalized by asking whether initial correlations between $A$ and a third reference system $R$ can propagate to $B$. 
\begin{lemma}\label{lemma:non-correlating}
	$(U,\omega_B)$ is a catalytic dilation if and only if for any additional system $R$ we have
	\begin{align}\label{eq:non-correlating}
			\tr_{A}(U \rho_{RA} \ox \omega_B U^\dagger) = \rho_R\ox \omega_B
	\end{align}
	for all states $\rho_{RA}$ on $RA$. 
	It suffices to check Eq.~\eqref{eq:non-correlating} for a single maximally entangled state $\rho_{RA}$. 
\end{lemma}
This Lemma follows directly from Proposition~1 of \cite{lie_catalytic_2021}, and we provide an alternative, diagrammatic proof in appendix \ref{app:non-corr}.
In fact, \cite{lie_catalytic_2021} also shows that $(U,\omega_B)$ must admit a catalytic dilation even with a weaker condition: $\omega_{B}$ in the RHS of Eq.~\eqref{eq:non-correlating} replaced by a potentially different state $\omega'_{B}$ that may depend on $\rho_{RA}$.
We emphasize that Lemma~\ref{lemma:non-correlating} does not imply that the resulting channel on $A$ is unitary, because we do not assume that $\omega_B$ is pure, as in the information-disturbance tradeoff \cite{kretschmann_information-disturbance_2008}.

The study of catalytic channels has recently been motivated as the only way to achieve catalytic advantage which is robust in the presence of errors~\cite{son2024robust}. Typically, we say that $\rho$ can be catalytically converted to $\sigma$ by a unitary $U$ if there exists a state $\omega_B$ such that
\begin{align}
	\tr_B(U \rho\ox\omega_B U^\dagger) = \sigma,\quad \tr_A(U \rho\ox \omega_B U^\dagger) = \omega_B.
\end{align}
If $\rho$ and $\sigma$ are not unitarily equivalent, such a conversion is possible with a suitable catalyst $\omega_B$ if and only if $H(\sigma) > H(\rho)$ \cite{boes_catalytic_2018,wilming_entropy_2021,wilming_correlations_2022}. 
In general, $\omega_B$ is fine-tuned with respect to $\rho$ for it to be preserved exactly; however, small perturbations in the initial state $\rho$ will lead to small perturbations in the final state on $B$.
Such errors may accumulate in the catalyst upon reuse, eventually degrading it and rendering it useless.
It is therefore desirable to have \emph{robust catalysis}, where $\omega_B$ must be preserved exactly, at least for small perturbations of $\rho$. 
However, as observed in \cite{son2024robust}, if $\rho$ has full rank, any tolerance towards a non-zero state preparation error $\epsilon >0$ immediately implies that $\omega_B$ must be preserved exactly for \emph{all} input states due to linearity.

As discussed in the introduction, catalytic channels are always doubly-stochastic and can only increase entropy. This prompts a natural question: how do such channels fit into the broader class of doubly-stochastic quantum channels?
A rich structure emerges from this question, which we explore in the remainder of this section.
We consider four subclasses of doubly-stochastic (DS) quantum channels:
\begin{enumerate}[leftmargin=*,itemsep=0pt]
	\item Mixed unitary channels (MU),
	\item Catalytic channels (CAT),
	\item Doubly-stochastic equilibrating channels (EQ $\cap$ DS),
	\item Factorizable channels (F).
\end{enumerate}
We assume that the catalytic and equilibrating dilating system is finite-dimensional.
These classes of doubly-stochastic quantum channels now form a hierarchy:
\begin{align}\label{eq:strict_hierarchy}
	\text{MU} \subsetneq \text{CAT} \subsetneq \text{EQ} \cap \text{DS} \subsetneq \text{F} \subsetneq \text{DS},
\end{align}
wherein all inclusions are strict. 
In particular, the strict inclusions $\text{MU} \subsetneq \text{CAT} \subsetneq \text{EQ}$ were previously open problems \cite{lie_randomness_2021, lie_catalytic_2021} and are now proven in this work. 

We have already introduced the first three classes of channels.
Among them, the smallest set (mixed unitary channels) coincides with the other two sets when the environment state of the dilation is restricted. 
\begin{lemma}
	For a doubly-stochastic quantum channel $T$, the following are equivalent:
	\begin{enumerate}[itemsep=0pt, leftmargin = *]
		\item $T$ is a mixed unitary channel.
		\item $T$ admits a non-degenerate equilibrating dilation.
		\item $T$ admits a non-degenerate catalytic dilation.
	\end{enumerate}
\end{lemma}
\begin{proof}
$3\to2$ follows from the definition of the catalytic dilation. 
$1\to3$ can be proved using Eq.~\eqref{eq: MU with controlled unitaries}, because we can always split up probabilities into distinct summands and distribute them over more dimensions to make the catalytic dilation non-degenerate. 
It thus suffices to show $2\to1$.
Suppose that a doubly-stochastic channel $T$ admits a non-degenerate equilibrating dilation $(U,\omega_B)$.
From Proposition \ref{prop:FP-dilations} we find $[U,\1\ox \omega_B]=0$, and since $\omega_B$ is non-degenerate it follows that $[U,\1\ox \proj{i}] = 0$ for all $i$, where $\ket{i}$ denote the eigenvectors of $\omega_B$ to the (all distinct) eigenvalues $p_i$. Hence $U = \sum_i U_i\ox \proj{i}$ and $T(\rho) = \sum_i p_i U_i \rho U_i^\dagger$.
\end{proof}
We now discuss factorizable channels. 
We call a quantum channel \emph{exactly factorizable} if it is of the form
\begin{align}
	T(\rho) = \tr_{B}\left(U \rho \ox \frac{\1}{d_B} U^\dagger\right),
\end{align}
and \emph{strongly factorizable} if it is of the form $T(\rho) = \tr_B \left(U \rho \ox \omega_B U^\dagger\right)$
with $[U,\1\ox \omega_B]=0$. Strongly factorizable quantum channels correspond to convex mixtures of exactly factorizable channels and can be approximated arbitrarily well by exactly factorizable ones \cite{Shor2010}.
It follows immediately from Proposition \ref{prop:FP-dilations} that strongly factorizable quantum channels precisely correspond to 
doubly-stochastic quantum channels that admit an equilibrating dilation:
\begin{align}
	\text{EQ} \cap \text{DS} = \text{strongly factorizable}. 
\end{align}
From the perspective of the resource theory of thermodynamics \cite{Horodecki2013,Brandao2015} and informational non-equilibrium \cite{Gour2015}, exactly factorizable channels can also be seen as \emph{noisy operations}, corresponding to thermal operations where the system Hamiltonian is fully degenerate.
Then strongly factorizable maps correspond to a random choice of noisy operation. 

Both exactly factorizable and strongly factorizable maps are examples of the more general class of \emph{factorizable} quantum channels \cite{haagerup_factorization_2010}, where $\omega_B$ is a tracial state on an arbitrary finite von Neumann algebra $\M$ and $U \in \B(\H_A)\ox \M$.
It is known that the set of factorizable maps on an $n$-dimensional Hilbert space coincides with the set of strongly factorizable maps for all $n$ if and only if the \emph{Connes embedding problem} ~\cite{connes_classification_1976} has an affirmative answer \cite{haagerup_factorization_2010, haagerup_asymptotic_2015}.
The recent $\text{MIP}^*=\text{RE}$ result \cite{ji_mipre_2022} claims that the Connes embedding problem has a negative answer.
If true, then there exists a finite dimension $n$ and a factorizable map on $n$-dimensional quantum systems that cannot be approximated by exactly factorizable maps~\footnote{At the time of writing, no completely peer-reviewed proof of \cite{ji_mipre_2022} has been published.}. We thus find:
\begin{align}
	\text{EQ} \cap \text{DS} \subsetneq \text{F}.
\end{align}
Moreover, it has been shown in \cite{haagerup_factorization_2010} that not all doubly-stochastic channels are factorizable, giving $\text{F} \subsetneq \text{DS}$.

The proofs for strict inclusions $\text{MU} \subsetneq \text{CAT} \subsetneq \text{EQ}\cap\text{DS}$ are shown in Appendix \ref{app:factorizable}.
We show $\text{CAT} \subsetneq \text{EQ}\cap \text{DS}$ by first showing that catalytic channels that admit a maximally mixed catalytic dilation (hence are exactly factorizable) can be extremal among the doubly-stochastic channels only if they are unitary. 
But the existence of non-unitary, exactly factorizable channels that are extremal among the doubly-stochastic maps is known \cite{haagerup_extreme_2021}. For $\text{MU} \subsetneq \text{CAT}$, we make use of \emph{Schur multiplier} channels, which are defined relative to some fixed basis and act as $T(\rho) = \rho\circ X$, where $X$ is a positive semidefinite matrix with unit diagonal entries and $\circ$ is the Schur-product, acting by componentwise multiplication. 
We show that if $X$ has real entries, then $T$ admits a maximally mixed catalytic dilation. However, there are examples of Schur multipliers with real $X$ which are not mixed unitary channels \cite{haagerup_factorization_2010}.

\textit{Conclusions.---}%
We start from a simple notion of information theoretic equilibrium, and demonstrate that local Hamiltonians naturally arise, along with the well-studied class of thermal operations. 
These operations are precisely characterized by their admission of equilibrating dilations, ensuring the absence of non-equilibrium resources. Our findings establish a clear distinction between thermal operations and its supersets of interest, such as enhanced thermal operations or Gibbs-preserving maps. 
We also use the notion of multi-partite equilibrium to show that thermal operations cannot have a robust version of catalytic advantage (Lemma \ref{lem:RCTO=TO}, End Matter). This resolves an open question from \cite{son2024robust}, and is particularly notable because it privileges thermal operations over its subsets of interest.  While thermal operations fully capture robust catalytic advantages through the use of passive heat baths/thermal environments, their subsets are more restrictive and thus fail to encompass all such advantages. 
Specifically, thermal processes such as elementary thermal operations~\cite{Lostaglio2018ETO} and Markovian thermal operations~\cite{Lostaglio2022MTO} still require additional, explicit use of robust catalysts, in order to expand their achievable set of operations~\cite{son2024catalysis,son2024hierarchy}. Together, these results underscore the significance of thermal operations, highlighting them as an operationally orthodox yet sufficiently general framework that naturally encompasses thermalization processes.

Drawing inspiration from the notion of heat baths, we introduced catalytic channels, a special class of doubly-stochastic quantum channels.
Within this broader class, we examined the relationships between mixed, catalytic, equilibrating, and factorizable quantum channels -- demonstrating that they are all distinct. Our results contribute to the broader question of what uniquely distinguishes mixed unitary channels from general doubly-stochastic quantum channels, highlighting the rich mathematical structure of doubly-stochastic channels, which is absent in the classical regime.


\emph{Acknowledgements.}  SHL is supported by the 2025 Research Fund (1.250007.01) of Ulsan National Institute of Science \& Technology (UNIST) and Institute of Information \& Communications Technology Planning \& Evaluation (IITP) Grant (RS-2023-00227854, RS-2025-02283189, ``Quantum Metrology Theory Based on Temporal Correlation Operators"). JS and NN are supported by the start-up grant of the Nanyang Assistant Professorship at the Nanyang Technological University in Singapore. 

\bibliographystyle{apsrev4-1}
\bibliography{newrefs.bib}

\appendix
\section*{End Matter}
\textit{Proofs of main results.---}%
Hereby we present the proofs of Lemma \ref{lemma:basic} and Proposition \ref{prop:FP-dilations}.
\begin{proof} (Lemma \ref{lemma:basic})
	The von Neumann entropy $H(\omega) := -\tr(\omega \log(\omega))$ is unitarily invariant, additive over tensor-products and fulfills $H(\omega_{AB}) = H(\omega_A) + H(\omega_B) - I(A:B)_\omega$, where $I$ denotes the mutual information. 
    From unitary invariance, additivity and the equilibrium condition we have $H(\omega_A) + H(\omega_B) = H(U\omega_A \ox \omega_BU^\dagger ) = H(\omega_A) + H(\omega_B) - I(A:B)_{U\omega_A\otimes\omega_B U^\dagger}$. 
    It is well-known that $I(A:B)_\omega =0$ if and only if $\omega_{AB} = \omega_A \ox \omega_B$. 
\end{proof}

\begin{proof} (Proposition \ref{prop:FP-dilations})
	Item \ref{item:FP-dilations1} is a restatement of Lemma~\ref{lemma:basic}. Item~\ref{item:FP-dilations2} follows from Item \ref{item:FP-dilations1}, since
	\begin{align}
		&T(\omega_A^{it} \rho_A \omega_A^{-it})\nonumber \\&= \tr_B(U(\omega_A\ox \omega_B)^{it}(\rho_A\ox\omega_B)(\omega_A\ox\omega_B)^{-it}U^\dagger) \nonumber \\
		&= \tr_B((\omega_A\ox \omega_B)^{it}U(\rho_A\ox\omega_B)U^\dagger(\omega_A\ox\omega_B)^{-it})\nonumber \\
		&= \omega_A^{it} T(\rho_A)\omega_A^{-it}.
	\end{align}
	Since $[U,\omega_A\ox\omega_B]=0$, $U$ acts unitarily on the supporting subspace of $\omega_A \ox \omega_B$. Hence Item~\ref{item:FP-dilations3} follows by restricting the Hilbert-space of $B$ to the support of $\omega_B$.
\end{proof}

\textit{Equilibrium for multiple subsystems.---}%
So far, we discussed equilibrium and its variations only in bipartite settings.
In this section, we extend this to multi-partite scenarios.
Let us start from the tripartite setting 
\begin{align}\label{eq:tri-partite equilibrium}
	\Tr_{\setminus X}(U \omega_{A}\otimes\omega_{B}\otimes\omega_{C} U^{\dagger}) = \omega_{X},
\end{align}
where $\Tr_{\setminus X}$ stands for the partial trace over all subsystems except for $X$ for each $X = A,B,C$. 
\begin{lemma}[Basic Lemma for multi-partite cases]\label{lemma: basic lemma multipartite}
	If $(\omega_{A},\omega_{B},\omega_{C})$ are in equilibrium under $U$ as in Eq.~\eqref{eq:tri-partite equilibrium}, 
	\begin{align}
		U \omega_{A}\otimes\omega_{B}\otimes\omega_{C} U^{\dagger} = \omega_{A}\otimes\omega_{B}\otimes\omega_{C}.
	\end{align}
\end{lemma}
Note that this Lemma does not directly follow from the Basic Lemma~\ref{lemma:basic}, since it is not clear whether this dilation is equilibrating with respect to the partition $AB|C$.
\begin{proof}
	First define $\omega_{ABC} = U\omega_{A}\otimes\omega_{B}\otimes\omega_{C}U^{\dagger}$, and $\omega_{AB} = \Tr_{C}(\omega_{ABC})$.
	Then, using the entropy argument and the fact that $\Tr_{AB}(\omega_{ABC}) = \omega_{C}$, 
	\begin{align}
		&H(\omega_{A})+H(\omega_{B})+H(\omega_{C})\nonumber\\
		 = &H(\omega_{AB})+H(\omega_{C}) - I(AB:C)_{\omega_{ABC}}.
	\end{align}
	Further writing $H(\omega_{AB}) = H(\omega_{A})+H(\omega_{B}) - I(A:B)_{\omega_{AB}}$, we obtain $I(A:B)_{\omega_{AB}} = 0$ and $I(AB:C)_{\omega_{ABC}} = 0$. 
	The former implies $\omega_{AB} = \omega_{A}\otimes\omega_{B}$, which combined with the latter proves the theorem. 
\end{proof}
It is clear that this lemma can immediately be generalized to any multi-partite settings. 
Now, we apply this multi-partite lemma to settings that extend beyond the catalytic dilations defined with unitaries, as in Definition~\ref{def:catalytic}, to incorporate more general quantum channels or processes for the dilation. 
A natural extension is the \emph{catalytic channel} of the form 
\begin{align}\label{eq:robust_catalytic_TO}
	T(\rho_{A}) = \tr_{C} (\mathcal{E}(\rho_{A}\otimes\omega_C)), 
\end{align}
with the constraint that for any input state $\rho_{A}$, the $C$ marginal state $\tr_{A} (\mathcal{E}(\rho_A\otimes\omega_C)) = \omega_C$.
Similarly to the earlier case of unitary dilation, this condition is equivalent to robust catalysis, where the catalyst remains invariant under small but arbitrary errors in $\rho_{S}$ \cite{son2024robust}.
If $\mathcal{E}$ in the dilation can be chosen arbitrarily, a trivial choice $\mathcal{E} = T \otimes \mathrm{id}_{C}$ with the identity channel $\mathrm{id}_{C}$ can be made. 
However, within the framework of resource theories, $\mathcal{E}$ must be chosen from a fixed set of free operations $\mathcal{F}$ defined by physical considerations.
This restriction then limits the set of catalytic channels.
For instance, catalytic channels defined as in Definition~\ref{def:catalytic} are restricted to be a subset of doubly-stochastic channels because $\mathcal{E}$ is assumed to be unitary. 

In \cite{son2024robust}, a wide range of resource theories are classified into those that admit catalytic channels outside the set of free operations and those that do not. 
Yet, the technique used in this classification is limited to sets of free operations that are completely resource non-generating (CRNG) \cite{Chitambar2019Review}, leaving the robust catalytic advantage for more general free operations as a largely open problem.
In this work, we show that whenever the set of free operations is defined as those with an equilibrating dilation (a condition that does not imply CRNG), the corresponding catalytic channels also admit equilibrating dilations.
In particular, this implies that for thermal operations, robust catalysis does not provide any advantage.  
\begin{lemma}\label{lem:RCTO=TO}
	All robust catalytic thermal operations can be implemented simply with thermal operations without a catalyst. 
\end{lemma}
\begin{proof}
	Suppose $\mathcal{B}$ is a thermal operation on $AC$ with dilation 
	\begin{align}
		\mathcal{B}(\varrho_{AC}) = \Tr_{B}(U\varrho_{AC}\otimes\omega_{B} U^{\dagger}),
	\end{align}
	and the energy-preserving unitary $[U,\omega_{A}\otimes\omega_{C}\otimes\omega_{B}] = 0$ for Gibbs states $\omega_{A},\omega_{C},\omega_{B}$.
	Hence, $\mathcal{B}$ is Gibbs-preserving, i.e. $\mathcal{B}(\omega_{AC}) = \omega_{AC}$, where we denote $\omega_{AC} = \omega_{A}\otimes\omega_{C}$.
	Then, a robust catalytic thermal operation is a channel $T(\rho_{A}) = \Tr_{C}(\mathcal{B}(\rho_{A}\otimes\tau_{C}))$ such that 
	\begin{align}\label{eq:TO_with_RC}
		\Tr_{AB}(U\rho_{A}\otimes\tau_{C}\otimes\omega_{B} U^{\dagger}) = \tau_{C},
	\end{align}
	with some catalyst state $\tau_{C}$ and for all system state $\rho_{A}$.

	We first show that the channel $T$ is Gibbs-preserving, i.e. 
	\begin{align}\label{eq:TO_with_RC2}
		\Tr_{CB}(U\omega_{A}\otimes\tau_{C}\otimes\omega_{B} U^{\dagger}) = \omega_{A},
	\end{align}	
	The monotonicity of the quantum relative entropy gives $D(\varrho_{AC}\| \omega_{AC}) \geq D(\mathcal{B}(\varrho_{AC})\|\omega_{AC})$ for any state $\varrho_{AC}$ and any thermal operation $\mathcal{B}$.
	Putting $\varrho_{AC} = \omega_{A}\otimes\tau_{C}$, we get $D(\omega_{A}\otimes\tau_{C}\| \omega_{AC}) = D(\tau_{C}\|\omega_{C})$ for the LHS from $\omega_{AC} = \omega_{A}\otimes\omega_{C}$.
	For the RHS, $D(\mathcal{B}(\omega_{A}\otimes\tau_{C})\|\omega_{AC}) \geq D(T(\omega_{A})\|\omega_{A}) + D(\tau_{C}\|\omega_{C})$, from super-additivity of the quantum relative entropy and Eq.~\eqref{eq:TO_with_RC}. 
	Since $D(T(\omega_{A})\|\omega_{A}) \geq 0$ with equality if and only if $T(\omega_{A}) = \omega_{A}$, the channel $T$ must be Gibbs-preserving.	
	Finally, we derive
	\begin{align}\label{eq:TO_with_RC3}
		\Tr_{AC}(U\omega_{A}\otimes\tau_{C}\otimes\omega_{B} U^{\dagger}) = \omega_{B},
	\end{align}	
	again using the same argument for Eq.~\eqref{eq:TO_with_RC2} with $\mathcal{B}$ replaced by $\mathcal{B}'(\varrho_{CB}) = \Tr_{A}(U\omega_{A}\otimes\varrho_{CB} U^{\dagger})$ and $\varrho_{AC}$ replaced by $\varrho_{CB} = \tau_{C}\otimes \omega_{B}$.
	
	Eqs.~\eqref{eq:TO_with_RC}--\eqref{eq:TO_with_RC3}, when combined, demonstrates that $(\omega_{A},\tau_{C},\omega_{B})$ are in equilibrium under $U$. 
	Lemma~\ref{lemma: basic lemma multipartite} then implies $[U,\omega_{A}\otimes\tau_{C}\otimes\omega_{B}] = 0$. 
	Restricting to the supporting subspace of $\tau_C$ and interpreting $\tau_C$ as a thermal state of some Hamiltonian, the channel $T$ has the usual dilation of a thermal operation
	\begin{align}
		T(\rho_{A}) = \Tr_{CB}(U \rho_{A}\otimes \omega_{CB}U^{\dagger}),
	\end{align}
	with the Gibbs state $\omega_{CB} = \tau_{C}\otimes\omega_{B}$.
\end{proof}


\clearpage
\onecolumngrid
\appendix

\section{Non-correlating versus catalytic dilations}\label{app:non-corr}

\begin{definition}[Non-correlating dilation]\label{def:nc_omega}
	We call a dilation $(U,\sigma_\sC)$ of a quantum channel \emph{non-correlating with respect to $\omega_\sS$} if
	\begin{align}\label{eq:nc-wrt-omega}
		\tr_{\sS}[(\1_{\bar A}\ox U)\proj{\omega}_{\bar\sS\sS} \ox \sigma_\sC(\1_{\bar A}\ox U^\dagger)] = \tr_{\sS}[\proj\omega_{\bar\sS\sS}]\ox \tau_\sC
	\end{align}
	for some purification $\ket\omega_{\bar\sS\sS}$ of $\omega_\sS$ on $\bar\sS\sS$ and some state $\tau_\sC$ on $\sC$.
\end{definition}
We have $\ket\omega_{\bar\sS\sS} = \sqrt{d_\sS}(W \sqrt{\omega}W^\dagger) \ox\1 \ket\Omega_{\bar \sS\sS}$, where $W: P_\omega \H_\sS \rightarrow \H_{\bar\sS}$ is some isometry and $P_\omega$ the support projection of $\omega_\sS$. It follows that the above definition does not depend on the choice of purification.
Moreover, Definition~\ref{def:nc_omega} is equivalent to a seemingly stronger condition, namely
\begin{align}
        \tr_\sS((\1_{\bar A}\ox U) \rho_{\bar\sS\sS} \ox\sigma_\sC (\1_{\bar A}\ox U^{\dagger})) = \rho_{\bar\sS}\ox \tau_\sC,
	\end{align}
for any system $\bar\sS$ and any initial state $\rho_{\bar \sS\sS}$ on $\bar\sS\sS$.

In \cite{lie_catalytic_2021} catalytic dilations of quantum channels where characterized through properties of the dilating unitary $U$ on $\sS\sC$.
To state this characterization we denote by ${}^{\top_\sS}$ the \emph{partial transpose}, i.e., the linear map defined by
\begin{align}
	(X_{\sS} \otimes Y_{\sC})^{\top_\sS} = X_{\sS}^\top \otimes Y_{\sC},
\end{align}
where $\top$ denotes the transpose.

One of the main technical tools used in this work is the relationship between catalytic dilations and non-correlating dilations, and the properties of the dilating unitary. We detail this as Proposition \ref{thm:partialtranspose} below.

\begin{proposition}[\cite{lie_catalytic_2021}]\label{thm:partialtranspose} Let $U$ be a unitary on $\sS \sC$ and $\sigma_\sC$ a state on $\sC$. The following are equivalent:
	\begin{enumerate}[itemsep=0pt,leftmargin= *]
\item\label{item:partialtranspose} $U^{\top_\sS}$ is unitary and $U(\1\ox \sigma_\sC) U^\dagger = \1\ox (W^\dagger\sigma_\sC W)$ for a unitary $W$ on $\sC$.
		\item\label{item:cat} $((1\ox W)U,\sigma_\sC)$ is a catalytic dilation.
		\item\label{item:nc} $(U,\sigma_\sC)$ is a non-correlating dilation.	
	\end{enumerate}
\end{proposition}
\noindent In particular, if $U^{\top_\sS}$ is unitary, $\sigma_\sC$ yields a catalytic dilation if and only if $[U,\1\ox\sigma_\sC]=0$.\\

We here provide a diagrammatic proof of Proposition \ref{thm:partialtranspose} for completeness.
Let us first recall a simple observation about the partial transpose. 
Let $U$ be a unitary on $\sS\sC$ and $\sS_2$ and $\sS_3$ be copies of $\sS$, i.e., there are unitary operators $W_j:\H_\sS\rightarrow \H_{\sS_j}$.
Associated are maximally entangled states $\ket{\Omega_2}_{\sS\sS_2},\ket{\Omega_3}_{\sS\sS_3}$ defined through
\begin{align}
	\bra{\Omega_j} (\ket i_{\sS}\ox \1_{\sS_j}) = \frac{1}{\sqrt{d_\sS}}\bra{i}_{\sS}W_j^\dagger
\end{align}
for some basis $\{\ket i_\sS\}$ on $\sS$.
With a suitable implicit reordering of tensor factors, we have
\begin{align}
	d_\sS (\bra i_\sS \ox \1_\sC)& (\bra{\Omega_3}_{\sS\sS_3} \ox \1_{\sS_2}\ox \1_\sC) (W_2^\dagger\ox W_3\ox U)(\1_{\sS_3} \ox \ket{\Omega_2}_{\sS\sS_2}\ox \1_\sC)(\ket j_\sS \ox \1_\sC)\\
	&= d_\sS (\bra{\Omega_3}_{\sS\sS_3}(W_3\ket j_{A} \ox \1_{\sS_3})\ox \1_\sC) U ((W_{2}^{\dagger}\bra i_{\sS} \ox \1_{\sS_2})\ket{\Omega_2}_{\sS\sS_2})\ox \1_\sC) \\
	&= (\bra j_\sS \ox \1_\sC)U(\ket i_\sS \ox \1_\sC)\\
	&= (\bra i_\sS \ox \1_\sC) U^{\top_\sS} (\ket j_\sS \ox \1_\sC).\\
	\implies \quad U^{\top_\sS} &= d_\sS (\bra{\Omega_3}_{\sS\sS_3} \ox \1_{\sS_2}\ox \1_\sC) (W_2^\dagger\ox W_3\ox U)(\1_{\sS_3} \ox \ket{\Omega_2}_{\sS\sS_2}\ox \1_\sC),	
	\end{align}
This calculation takes a simple form in terms of the tensor-network diagram (to be read from bottom to top)
\begin{align}\label{diag:partial-transpose}
	d_{\sS}\ \vcenter{\hbox{\includegraphics{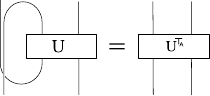}}}\phantom{1},
\end{align}
where half-circles correspond to maximally entangled states and the unitaries $W_j$ are suppressed.
We show the equivalence of three identities using the diagrams. 
The equivalence of the first two are given by linearity in $X$:
\begin{align}\label{eq:central-identity-1}
	\vcenter{\hbox{\includegraphics{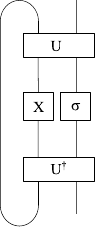}}}=	\vcenter{\hbox{\includegraphics{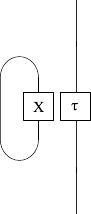}}}\ \forall X\ \Leftrightarrow 
	d_\sS^2 \ \vcenter{\hbox{\includegraphics{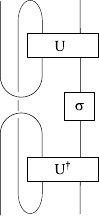}}}=	\vcenter{\hbox{\includegraphics{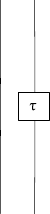}}},
\end{align}
and for the remaining one we use Eq.~\eqref{diag:partial-transpose} to show
\begin{align}\label{eq:central-identity-2}
	d_{\sS}^2	\vcenter{\hbox{\includegraphics{diagrams/central-identity-3.pdf}}}=	\vcenter{\hbox{\includegraphics{diagrams/central-identity-4.pdf}}}\quad\Leftrightarrow
	\vcenter{\hbox{\includegraphics{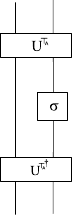}}}=	\vcenter{\hbox{\includegraphics{diagrams/central-identity-4.pdf}}}\,.
\end{align}
\begin{lemma}\label{lemma:partial-transpose}
	The pair $(U,\1/d_\sC)$ is a catalytic dilation of a quantum channel $T$ on $\sS$ if and only if $U^{\top_\sS}$ is unitary.
\end{lemma}
\begin{proof}
	$(U,\1/d_\sC)$ is a catalytic dilation if and only if the LHS of Eq.~\eqref{eq:central-identity-1} holds for $\tau=\sigma=\1/d_\sC$, because density matrices span all linear operators.
    On the other hand, $U^{\top_\sS}$ is unitary if and only if the RHS of Eq.~\eqref{eq:central-identity-2} is true for $\tau=\sigma=\1/d_\sC$.
    Hence, Eqs.~\eqref{eq:central-identity-1} and~\eqref{eq:central-identity-2} prove the claim. 
\end{proof}

\begin{lemma}\label{lemma:nc-doubly-stochastic}
	Let $T$ be a quantum channel on $\sS$ that admits a non-correlating dilation $(U,\sigma_\sC)$. Then the entropy on $\sC$ is invariant if the input state on $\sS$ is maximally mixed. 
\end{lemma}
\begin{proof}
	The purification of the maximally mixed input state on $\sS$ is a maximally entangled initial state $\rho_{\bar\sS \sS} =\proj\Omega_{\bar\sS\sS}$.
	To simplify notation, in the following we use primes to denote subsystems after the application of the unitary $\1\ox U$ while unprimed systems refer to the initial state $\rho_{\bar\sS\sS}\ox\sigma_\sC$. 
	From the Araki-Lieb inequality~\cite{araki_entropy_1970} we have 
	\begin{align}
		S(\sC)&= S(\bar\sS \sS\sC) = S(\bar\sS' \sS'\sC') \geq | S(\bar \sS' \sC') -  S(\sS')| = |S(\bar\sS')+ S(\sC') - S(\sS')|,
	\end{align}
	where we used that the dilation is non-correlating in the last step. Since the initial state is maximally entangled and $U$ only acts on $\sS\sC$, we further have $S(\bar\sS')= S(\bar\sS)= S(\sS)$ and $S(\sS') \leq S(\sS)$.
	Thus $S(\sC)-S(\sC') \geq S(\sS) - S(\sS') \geq 0$.
	On the other hand, from  unitary invariance of von Neumann entropy we get
	\begin{align}
		0 &\leq I(\sS': \sC') = S(\sS') + S(\sC') - S(\sS'\sC') = S(\sS') +S(\sC') - S(\sS\sC) \\
		&= S(\sS') - S(\sS) + S(\sC') - S(\sC). 
	\end{align}
	Hence $S(\sC) - S(\sC') \leq -(S(\sS) - S(\sS'))\leq 0$ and therefore $S(\sC)=S(\sC')$. 
\end{proof}

\begin{proof}[Proof of Proposition \ref{thm:partialtranspose}] We show the following set of relations between the items of the proposition statement.
	\begin{itemize}[leftmargin= *]
		\item \ref{item:partialtranspose} $\Leftrightarrow$ \ref{item:cat}: 
		By adjusting $U$ we can assume without loss of generality that $W=\1$.
		"$\Rightarrow$" follows, because $[U,\1\ox\sigma_\sC]=0$ is equivalent to $[U^{\top_\sS},\1\ox\sigma_\sC]=0$. Hence the RHS of Eq.~\eqref{eq:central-identity-2} holds with $\tau_\sC=\sigma_\sC$, which proves the claim using Eq.~\eqref{eq:central-identity-1}.  
			For the converse, Proposition \ref{prop:FP-dilations} implies that $[U,\1\ox\sigma_\sC]=0$ because every catalytic channel is an equilibrating channel. Therefore $U = \oplus_i U_i$ relative to a decomposition $\H_\sS\ox\H_\sC = \oplus_i \H_\sS \ox \H_{\sC,i}$
		with $\sigma = \oplus_i(q_i \1_i/d_i)$. Thus each $(U_i,\1_i/d_i)$ is a catalytic dilation of a  channel $T_i$ with $T=\sum_i q_i T$. 
		Hence $U^{\top_\sS}_i$ is unitary by Lemma \ref{lemma:partial-transpose} and therefore $U^{\top_\sS} = \oplus U^{\top_\sS}_i$ is unitary.
		\item \ref{item:cat} $\Rightarrow$ \ref{item:nc}: The assumption \ref{item:cat} implies that the LHS of Eq.~\eqref{eq:central-identity-1} is true for $\tau = \sigma$. By multiplying the RHS of Eq.~\eqref{eq:central-identity-1} by $\rho_{\sS}^{1/2}\ox \1$ from above and below, we obtain \ref{item:nc}.
		\item \ref{item:nc} $\Rightarrow$ \ref{item:cat}: By assumption, the RHS of Eq.~\eqref{eq:central-identity-1} is satisfied for some density matrix $\tau_\sC$.
		The LHS together with $X = \1_\sS/d_\sS$ implies that $\sigma_{\sC} \succeq \tau_\sC$, since the induced quantum channel on $\sC$ is exactly factorizable and hence doubly-stochastic. 
		On the other hand, Lemma \ref{lemma:nc-doubly-stochastic} shows that $S(\sC)_\sigma = S(\sC)_\tau$. Since von Neumann entropy is strictly Schur-concave, this implies that  
		$W \tau_\sC W^\dagger  = \sigma_{\sC}$ for some unitary $W$. Thus $(1\ox W)U$ satisfies the RHS of Eq.~\eqref{eq:central-identity-1} with $\tau=\sigma$. Hence $\big((1\ox W)U,\sigma_\sC)$ is a catalytic dilation by the LHS. 
	\end{itemize}
\end{proof}

\section{Relation to dual-unitary circuits}
\label{app:dual-unitary}
In Appendix~\ref{app:non-corr} we have seen that catalytic channels are induced by unitary operators whose partial transpose is also unitary.
Once we introduce bases (an identification between vector space and dual space), a linear map $V: \H_1 \ox\H_2\to \H_3\ox \H_4$ can also be read as a linear map $V^\Gamma : \H_1 \ox \H_3 \to \H_2\ox \H_4$. In graphical tensor-network we simply write:
\begin{align}
V = \vcenter{\hbox{\includegraphics{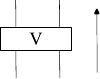}}} \quad \text{versus} \quad   V^\Gamma = \vcenter{\vspace{-.3cm}\hbox{\includegraphics{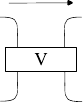}}},
\end{align}
where the arrow indicates the direction of the mapping.
A unitary operator $U$ is called \emph{dual-unitary} if $U^\Gamma$ is also unitary. Note that this requires $\H_1 \ox \H_3 \cong \H_2\ox \H_4$. 

It is easy to see diagrammatically that a unitary $U:\H_A\ox \H_B \to \H_A\ox \H_B$ is a catalytic unitary, i.e., its partial transpose is unitary, if and only if the unitary operator $U\mathbb S : \H_B\ox \H_A \to \H_A\ox \H_B$ is dual-unitary, where $\mathbb S$ is the swap operator:
\begin{align}
U^{\top_A} = \vcenter{\hbox{\includegraphics{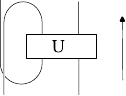}}} \quad \text{unitary} \ \Leftrightarrow  \quad   (U\mathbb S)^\Gamma = \vcenter{\vspace{.3cm}\hbox{\includegraphics{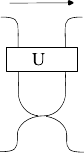}}}\quad \text{unitary}.
\end{align}
Thus there is a correspondence between catalytic unitaries and dual-unitary operators.

\section{MU $\subsetneq$ CAT $\subsetneq$ EQ}
\label{app:factorizable}
We prove in this appendix the strictness of the hierarchy presented in the main text (Eq.~\eqref{eq:strict_hierarchy}, or Fig.~\ref{fig:hierarchy}).
\subsection{CAT $\subsetneq$ EQ}
\begin{lemma}\label{lemma:extremal-unitary}
	Let $T$ admit an exact factorization that is also a catalytic dilation. 
	Then either $T$ is unitary, or $T$ is a non-trivial convex mixture of doubly-stochastic quantum channels. 
	In other words, $T$ is not extremal in the set of doubly-stochastic channels if it is not unitary.  
\end{lemma}
\begin{proof}
	By assumption of the lemma $T$ has a catalytic dilation $(U,\1/d_\sC)$. Choose some orthonormal basis $\{\ket j\}$ on $\sC$.
	We have
	\begin{align}
		T(\rho) = \sum_{j=1}^{d_\sC} \frac{1}{d_\sC}\tr_2(U \rho \ox \proj j U^\dagger) ) = \sum_j \frac{1}{d_\sC} T_{\ket j}(\rho),
	\end{align}
	with
	\begin{align}
		T_{\ket j}(\rho) = \tr_2(U \rho \ox \proj j U^\dagger) = \1\ox \bra j (U^{\top_\sC} \rho \ox \1 {U^{\top_\sC}}^\dagger ) \1\ox\ket j.
	\end{align}
	By Proposition \ref{thm:partialtranspose}, $U^{\top_\sS}$ is unitary, which implies that $U^{\top_\sC} = (U^{\top_\sS})^\top$ is also unitary.
	Hence 
	\begin{align}
		T_{\ket j}(\1) = \1 \ox \bra j (\1 \ox \1) \1\ox \ket j = \1,
	\end{align}
	i.e. each $T_{\ket{j}}$ is a doubly-stochastic quantum channel. 
	Since $T$ is a uniform mixture of $T_{\ket{j}}$, it is a non-trivial mixture of doubly-stochastic channels, unless $T_{\ket{j}} = T$ for all $j$.
	Repeating the argument for all possible orthonormal bases we find that either $T$ is a non-trivial mixture of doubly-stochastic channels or $T_{\ket \psi}=T$ for all normalized $\ket \psi\in \H_\sC$.
	Suppose the latter is true and write $A = U^{\top_\sS} \rho \ox \1 (U^{\top_\sS})^\dagger$. Then we have
	\begin{align}
		T(\rho) = (\1 \ox\bra \psi) A (\1\ox \ket\psi) 
	\end{align}
	for all $\psi$.	This implies $U^{\top_\sS} \rho \ox \1 (U^{\top_\sS})^\dagger = A = T(\rho)\ox \1$ for all $\rho$.
	Since $U^{\top_\sS}$ is unitary this is possible only when $ \rho $ and $T(\rho)$ have the same spectrum (including multiplicities), i.e. there exists a unitary $W$ such that $T(\rho) = W\rho W^\dagger$. 
\end{proof}

As pointed out in the main text, there exist non-unitary and exactly factorizable channels that are extremal among the doubly-stochastic maps \cite{haagerup_extreme_2021}. 
According to Lemma~\ref{lemma:extremal-unitary} these channels must lie outside CAT, showing a strict gap between CAT and EQ.

\subsection{MU $\subsetneq$ CAT}
Next, we detail some technical ingredients used to show the strict inclusion of mixed unitary channels in the set of catalytic dilations. To do so, we first need to introduce the notion of Schur multipliers. In the following, we denote by $M_n(\CC)$ ($M_n(\RR)$) the set of $n\times n$ matrices with complex (real) coefficients. 
\begin{definition}[Schur multiplier]\label{def:Schur_multiplier}
	Let $X\in M_n(\CC)$ be a positive semidefinite matrix with $X_{ii}=1$ for $i=1,\ldots,n$. We define the associated doubly-stochastic quantum channel acting on $M_n(\CC)$ as 
	\begin{align}
		T_X(x) = x\circ X,
	\end{align}
	where $\circ$ denotes the Schur product $(x\circ X)_{ij} = x_{ij} X_{ij}$. 
	The channel $T_X$ is called a \emph{Schur multiplier}.
\end{definition}
Let us observe for now that $X_{ij} = \bra i T_X(\ketbra{i}{j}) \ket j$. 
There is a close connection between Schur multipliers and factorizable maps. Specifically, it was shown that a Schur multiplier $T_X$ is factorizable, if and only if 
\begin{align}
	X_{ij} = \tau(u_i u_j^\dagger),
\end{align}
where $\tau$ is a (faithful, normal) tracial state on a finite von Neumann algebra $\M$ and $u_i\in \M$ are unitaries \cite{haagerup_factorization_2010}.
If $u_i$ are finite-dimensional matrices, $T_X$ is exactly factorizable. 
It has been shown that the Connes embedding problem is equivalent to showing that all matrices $X$ as above may be approximated using unitaries on a finite-dimensional matrix algebra \cite{dykema_matrices_2009}.

We next show that all \emph{real} positive semidefinite matrices with diagonal entries equal to $1$ can be represented using finite-dimensional unitaries, yielding a catalytic dilation $T_X$ in terms of an exact factorization. 

\begin{proposition}\label{prop:real-schur}
	Let $X\in M_n(\RR)$ be positive semidefinite and $X_{ii}=1$ for $i=1,\ldots,n$. Then there exists a collection of $n$ self-adjoint unitary matrices $\{u_i\}_{i=1}^n \subset M_{2^d}(\CC)$ with $d=\mathrm{rank}(X)$ such that:
	\begin{enumerate}[itemsep=0pt,leftmargin= *]
		\item\label{item:realB} $X_{ij} = 2^{-d} \tr(u_i  u_j)$
		\item\label{item:schur-dilation}  	The Schur multiplier $T_X$ is exactly factorizable as $T_X(\rho) = \tr_2(U \rho \ox \frac{\1}{2^d} U^\dagger)$, where $U = \sum_j \proj{j}\ox u_j$.
		\item\label{item:schur-catalytic}  The pair $(U, \1/2^d)$ is a catalytic dilation of $T_X$.
	\end{enumerate}
\end{proposition}
\begin{proof}
	The proof combines several observations in \cite{haagerup_factorization_2010}. 
	First, by \cite[Remark 2.7]{haagerup_factorization_2010} we can write $T_B(x) = \sum_{i=1}^d a_i x a_i$, where $a_i\in M_n(\RR)$ are real, diagonal matrices that are linearly independent and fulfill $\sum_{i=1}^d a_i^2 = \1$. 	We now follow the proof of \cite[Corollary 2.5]{haagerup_factorization_2010}. Consider fermionic creation/annihilation operators $f_i^\dagger,f_j$ with $i,j=1,\ldots,d$ as matrices in $M_{2^d}(\CC)$ and define
	\begin{align}
		v_i &= f_i + f_i^\dagger, \qquad
		U  = \sum_{i=1}^d a_i\ox v_i.
	\end{align}
	Since $v_i v_j + v_j v_i = 2\delta_{ij}\1$, we find that each $v_i$ is self-adjoint and unitary  and $\tr(v_i^\dagger v_j) = \tr(v_i v_j) = \delta_{ij} 2^d$. Moreover $U$ is self-adjoint and unitary:
	\begin{align}
		U^\dagger U = U^2 = \sum_{i,j=1}^d a_i a_j\ox v_i v_j = \frac{1}{2}\sum_{i,j=1}^d (a_ia_j + a_j a_i)\ox v_i v_j &= \frac{1}{2} \sum_{i,j=1}^d a_i a_j \ox (v_i v_j + v_j v_i)
		= \sum_{i=1}^d a_i^2 \ox \1 = \1\ox\1.
	\end{align}
	Now consider the completely positive map $x\mapsto \frac{1}{2^d}\tr_{2}(U (x\ox \1) U)$, where $\tr_2 = \mathrm{id}\ox \tr$ denotes the partial trace. We have
	\begin{align}
		\frac{1}{2^d}\tr_{2}(U (x\ox \1) U) &= \frac{1}{2^d}\sum_{i,j=1}^d a_i x a_j \tr(v_i^\dagger v_j) = \sum_{i,j=1}^d a_i x a_j \delta_{ij} = T_X(x),
	\end{align}
	Since the $a_i$ are diagonal, we can write $U = \sum_{i=1}^n \proj{i}\ox u_i$ and since $U$ is self-adjoint and unitary, so are the $u_i$. From Def.~\ref{def:Schur_multiplier} we note that $X_{ij} = \bra i T_X(\ketbra{i}{j}) \ket j$, hence it follows that
	\begin{align}
		X_{ij} = \frac{1}{2^d}\bra i \tr_{2}(U (\ketbra{i}{j}\ox \1) U)\ket j = \frac{1}{2^d} \tr(u_i u_j),
	\end{align}
	which shows Items 1 and 2 of the proposition statement. Now let $\rho \in M_n(\CC)$ be a density matrix. Since $u_i^2=\1$ we find
	\begin{align}
		&(\tr \ox \mathrm{id})(U(\rho \ox \frac{\1}{2^d}) U^\dagger) = \sum_{i,j=1}^n \tr(\proj i \rho \proj j) \frac{u_i u_j}{2^d} = \sum_{i=1}^n \rho_{ii} \frac{u_i^2}{2^d} = \tr(\rho) \frac{\1}{2^d} = \frac{\1}{2^d},
	\end{align}
	showing Item \ref{item:schur-catalytic} of the proposition statement.
\end{proof}

However, there is a known example of the matrix $X$ satisfying the condition of Proposition \ref{prop:real-schur}, while the corresponding Schur multiplier $T_{X}$ cannot be written as a mixed unitary; see Example~3.3 of \cite{haagerup_factorization_2010}.
This concludes our proof of MU $\subsetneq$ CAT.

\end{document}